\documentclass{aa}  

\usepackage{graphicx}
\usepackage{amsmath,siunitx}   
\usepackage{leftidx}
\usepackage{tablefootnote}
\usepackage[dvipsnames]{xcolor}
\usepackage{textcomp} 
\usepackage{longtable}
\usepackage{tabularx, ltablex, caption}
\usepackage{caption} 
\usepackage{placeins}
\usepackage{txfonts}

\begin{document}

   \title{The [NII] $205\, \mu m$ line emission from high-z SMGs and QSOs}

   \subtitle{}

   \author{Saimurali Kolupuri\inst{1,2}
          ,
          Roberto Decarli\inst{1}
          ,
          Roberto Neri\inst{3}
          ,
          Pierre Cox\inst{4}
          ,
          Carl Ferkinhoff\inst{5}
          ,
          Frank Bertoldi\inst{6}
          ,
          Axel Weiss\inst{7}
          ,
          Bram P. Venemans\inst{8}
          ,
          Dominik A. Riechers\inst{9}
          ,
          Emanuele Paolo Farina\inst{10}
          ,
          Fabian Walter\inst{11}
          }
          

             


   \date{}

\institute{
INAF -- Osservatorio di Astrofisica e Scienza dello Spazio di
Bologna, via Gobetti 93/3, I-40129, Bologna, Italy\\
\email{saimuralikolupuri@gmail.com}\and
 Indian Institute of Science Education and Research, Govt. ITI Building Engg. School Junction, Brahmapur, Odisha 760010, India\and
Institut de Radioastronomie Millimétrique (IRAM), 300 rue de la Piscine, 38400 Saint-Martin-d’Hères, France\and
Sorbonne Université, CNRS UMR 7095, Institut d’Astrophysique de Paris, 98bis bvd Arago, 75014 Paris, France\and
Department of Physics, Winona State University, Winona, MN 55987, USA\and
Argelander-Institute for Astronomy, University of Bonn, Auf dem Hügel 71, D-53121 Bonn, Germany\and
Max-Planck-Institut für Radioastronomie, Auf dem Hügel 69 D-53121 Bonn, Germany\and
Leiden Observatory, Leiden University, Niels Bohrweg 2, NL-2333 CA Leiden, Netherlands\and
I. Physikalisches Institut, Universität zu Köln, Zülpicher Strasse 77, 50937 Köln, Germany\and
Gemini Observatory, NSF’s NOIRLab, 670 N A’ohoku Place, Hilo, Hawai’i 96720, USA\and
Max-Planck Institut für Astronomie, Königstuhl 17, D-69117, Heidelberg, Germany
}

 
  \abstract
  {We present [NII] 205 $\mu$m fine structure line observations of three submillimeter galaxies (SMGs) and three quasar host galaxies at 4$\lesssim$z$\lesssim$6 using the Institut de radioastronomie millim\'etrique (IRAM)
 interferometer. The [NII] emission is detected in three sources, and we report detections of the underlying dust continuum emission in all sources. The observed [NII]-to-infrared luminosity ratio spans at least 0.5 dex for our sources. Comparing our estimates with sources detected in the [NII] 205 $\mu$m at similar redshifts shows that the overall [NII]-to-IR luminosity ratio spans over a dex in magnitude from L$_{[NII]}$/L$_{IR}$ $\sim$ 10$^{-4}$ -- 10$^{-5}$ and follows the trend of the so-called [NII] fine structure line deficit observed in (ultra)-luminous infrared galaxies in the local Universe. The [CII]-to-[NII] luminosity ratio is >10 for most of our sources, indicating that the bulk of the [CII] 158 $\mu$m line emission (\emph{f}([CII]$^{PDR}$) >75\%) arises from the neutral medium. From our analysis, we do not find significant differences in the [NII] 205 $\mu$m emission and the respective ratios between SMGs and QSOs, suggesting a negligible contribution to the boosting of [NII] 205 $\mu$m emission due to the active galactic nucleus (AGN) photoionization. Future investigations involving other fine structure lines and optical diagnostics will provide further insight into a suite of ionized medium properties and reveal the diversity between AGN and non-AGN environments.}


   \titlerunning{The [NII] $205\, \mu m$ line emission from high-z SMGs and QSOs}
   \authorrunning{Kolupuri et al}
   \maketitle
%

\begin{table*}[!htp]
    \caption{The sample of this study}
    \centering
    \begin{tabular}{c c c c c c c c c c}
    \hline
         Source &  R.A. & Decl. & z$_{ref}$ &  Transition& $\mu^a$ &L$_{[CII]}$ & L$_{IR}$ & L$_{FIR}$/M$_{H_2}$ & Ref\\
          & J2000.0 & J2000.0 &  & & & [$10^9$L$_{\odot}$]  &[$10^{13}$L$_{\odot}$]&(L$_{\odot}$/M$_{\odot})$&\\
         (1) & (2) & (3) & (4) & (5) & (6) & (7) & (8) & (9) & (10)\\
         \hline
         GN20 & 12:37:11.90 & 62:22:12.1 & 4.0554&CO(2-1)  & 1 & ... & 1.52 $\pm$ 0.11 & 82 $\pm$ 8 & 4$^b$,11\\
         \\
         ID141 & 14:24:13.93 & 02:23:04.8 & 4.243&CO(5-4),(4-3) & 5.8 & 10.6 $\pm$ 1.7 & 1.58 $\pm$ 0.12 & 230 $\pm$ 10& 5$^b$,10\\
         \\
         HDF850.1 & 12:36:52.02 & 62:12:26.0 & 5.1853&[CII] 158 $\mu$m& {\color{black}2.5} & {\color{black}1.1} & 0.20 $\pm$ 0.05 &  255 $\pm$ 40 & 7$^b$, {\color{black}17}\\
         \\
         PSSJ2322+1944 & 23:22:07.18 & 19:44:22.4 & 4.1199&CO(5-4),(4-3) & 5.3 & <  1.7 & 0.55 $\pm$ 0.10 & 132 $\pm$ 12 & 1$^b$,2,3,8\\
         \\
         J2054-0005 & 20:54:06.51 & -00:05:14.6 & 6.0389&[CII] 158 $\mu$m& 1 & 3.36 $\pm$ 0.12       & 0.90 $\pm$ 0.06 & 296 $\pm$ 171 &13,14$^b$,16   \\
         \\
         J2310+1855 & 23:10:39.00 & 18:55:19.9 & 6.0031&[CII] 158 $\mu$m& 1 & 8.31 $\pm$ 0.41 & 2.29 $\pm$ 0.04 & 240 $\pm$ 60 &6$^b$,9,12,15       \\
         \hline
         
    \end{tabular}
    \tablefoot{(1) Source name. (2) Right ascension (J2000.0). (3) Declination (J2000.0). (4) Redshift. (5) Atomic/Molecular transition used to derive redshift estimates. (6) Gravitational magnification factor. (7) [CII] 158 $\mu$m luminosity. (8) Infrared luminosity (8-1000 $\mu$m). (9) Far infrared-to-molecular gas mass ratio (computed from literature). (10) References for magnification factor, [CII] 158 $\mu$m luminosity, and molecular gas mass. {\color{black}The IR luminosities were estimated by fitting the dust spectral energy distribution (SED) with a modified blackbody (see Appendix B)}. Luminosities are corrected for gravitational magnification.\\
    \textbf{References.} 1: Cox et al. (2002), 2: Pety et al. (2004), 3: Solomon \& Vanden Bout (2005), 4: Carilli et al. (2011), 5: Cox et al. (2011), 6: Wang et al. (2013), 7: Neri et al. (2014), 8: Valentino et al. (2018), 9: Shao et al. (2019), 10: Cheng et al. (2020), 11: Cortzen et al. (2020), 12: Li et al. (2020), 13: Pensabene et al. (2020), 14: Venemans et al. (2020) 15: Tripodi et al. (2022), 16: Salak et al. (2024), 17: Sun et al. (2024). \\
    $^a$: The magnification factors are based on measurements from either the [CII] 158 $\mu$m or the CO transitions. For this study, we will assume that the magnification strength is the same for the [NII] 205 $\mu$m emission as well.\\
    $^b$: References for redshift}
    \label{tab:my_label}
\end{table*}

\section{Introduction}

   Investigating the physical properties of the interstellar medium (ISM), such as the amount of molecular gas, dust opacity, electron density, hardness of radiation fields, and metal enrichment, can shed light on vital processes like the baryon cycle or the impact of the active
galactic nucleus (AGN) in galaxies. In local galaxies, these physical properties are typically estimated using rest frame optical/UV line diagnostics, for example, H$\alpha$, H$\beta$, [OII] 3727 \AA, [OIII] 5007 \AA, and [NII] 6584 \AA\, \citep[e.g.,][]{Baldwin1981, Veilleux1987, Keenan1996, Izotov2006}. However, these lines are either generally faint or inaccessible by ground-based telescopes at high redshift as they are shifted to wavelengths where the atmosphere is opaque and can only be targeted by sensitive space telescopes such as the James Webb Space Telescope \citep[JWST, ][]{Cameron2023, Katz2023, Sanders2023}. {\color{black}Excitingly, JWST has been successful in securing key optical diagnostics, such as the H$\alpha$ and [OIII] lines, from the very first galaxies at z$\geq$10 \citep{Alvarez-Marquez2024, Hsiao2024, Zavala2024}.}

   Prior to the {\color{black}historic} launch of JWST, however, such properties of high-redshift galaxies were studied by observing molecular or atomic tracers whose emission lines are shifted in the millimeter and submillimeter wavelength ranges and hence are accessible by ground-based telescopes, with an additional advantage of not being too much affected by dust obscuration. In particular, rotational molecular transitions, associated with molecules such as CO and H$_2$O, and forbidden atomic fine-structure lines emitted in the infrared (IR) from carbon, oxygen, nitrogen, and their ions (e.g., [CI] 370 $\mu$m, [CII] 158 $\mu$m, [OIII] 88 $\mu$m, [NII] 122 $\mu$m), have been detected in numerous high-redshift galaxies in the past few years \citep[e.g.,][]{Bertoldi2003, Walter2012, Carilli2013, Valentino2018, Vishwas2018, Riechers2019, Rybak2020, Harikane2020, Pensabene2021, Decarli2012, Decarli2014, Decarli2022, Yang2023}. A combination of these emission line diagnostics and their luminosity ratios can shed light on the various phases of the ISM and its properties, for example, opacity, electron density, star formation rates (SFRs), gas mass, metallicity, radiation field intensity, ionization strength, and excitation mechanisms \citep{De Looze2014, Herrera-Camus2015, Madden2020, Lamarche2022, Vizgan2022, Decarli2012, Decarli2023}. 

    While the [CII] 158 $\mu$m is the dominant and best-studied cooling line of the interstellar medium \citep{Carilli2015, Neeleman2017, Decarli2018, Lagache2018, Neeleman2019, Bethermin2020, Khusanova2022}, another vital fine structure line that studies have been increasingly focusing on is the [NII] 205 $\mu$m emission line \citep{Walter2009, Decarli2012, Decarli2014, Pavesi2016, Zhao2013, Zhao2016}. Arising from the $\leftidx{^3}{p}{_1} \rightarrow \leftidx{^3}{p}{_0}$ forbidden transition, this line is of particular interest for the following reasons: 1) with an ionization potential of 14.53 eV, it traces the bulk of the warm ionized medium; 2) in combination with other fine structure lines such as [CII] 158 $\mu$m, [NII] 122 $\mu$m, it enables constraints on key physical properties, such as the fraction of the [CII] 158 $\mu$m arising from the ionized medium,  metal enrichment, and electron density {\color{black}\citep[e.g.,][]{Cunningham2020, Doherty2020, Tadaki2022}}; 3) It can be used as an indicator for star formation rates (SFR) as both computation models \citep{Orsi2014} and observations \citep{Farrah2013, Zhao2013, Zhao2016} have shown a linear correlation between the [NII] 205 $\mu$m line luminosity and SFR in local star-forming and ultra luminous infrared galaxies (ULIRGs), and a good correlation with the star formation rate surface density \citep{Herrera-Camus2016}. While these studies show a good correlation between star formation and the [NII] 205 $\mu$m luminosity, the scatter is, however, plagued by, for example, the [NII] 205 fine structure line deficit. The choice of SFR-[NII] scaling relation depends on the prior knowledge of different factors, for instance, the far-infrared (FIR) color \citep{Zhao2016} or physical parameters such as electron density, nitrogen abundance, and ionization parameter \citep{Herrera-Camus2016}. A suitable sample has been produced for statistical analysis on local galaxies thanks to the advent of the Herschel Space Observatory \citep[e.g.,][]{Zhao2016}, but only a few sources have been detected at high redshifts. However, with the unparalleled sensitivities reached by the {\color{black} IRAM (Institut de RadioAstronomie Millimétrique) NOrthern Extended Millimetre Array (NOEMA)} and the {\color{black}Atacama Large Millimeter/submillimeter Array (ALMA)}, the number of galaxies detected in [NII] 205 $\mu$m has been increasing over the years \citep[e.g.,][]{Bethermin2016, Lu2018, Doherty2020, Schreiber2021, Meyer2022}.

    In this paper, we present [NII] 205 ${\mu }$m fine-structure line observations of three submillimeter galaxies (SMGs; GN20, ID141, HDF850.1) and three quasar host galaxies (QSOs; PSSJ2322+1944, SDSS J2054-0005, SDSS J2310+1855) at 4$\lesssim$z$\lesssim$6, that were carried out over a decade {\color{black}using the IRAM interferometer (both the Plateau de Bure Interferometer (PdBI) and, later, NOEMA)}. The paper is structured as follows: in Sect. 2, we introduce the selected sources and summarise the observations and the data reduction. In Sect. 3, we present and discuss the results, focusing on the {\color{black}[NII]-to-IR \& [CII]-to-[NII] ratios}, and, finally, in Sect. 4, we outline the main conclusions of this study. Throughout this paper, we assume a concordance cos\-mo\-lo\-gy with H$_0$=70\,km\,s$^{-1}$\,Mpc$^{-1}$, $\Omega_{\rm M}$=0.3, and \mbox{$\Omega_\Lambda$=1-$\Omega_{\rm M}$=0.7}.






\begin{table*}[!htp]
    \centering
    \caption{Observation parameters }
    \begin{tabular}{c c c c c c}
    \hline
         Source&Exp time& Beam & Beam PA & Line sensitivity & Continuum sensitivity \\
         &[hr] & [$''$] & [deg] & [mJy/beam] & [mJy/beam]\\
         (1)&(2)& (3)& (4) & (5) & (6)\\
         \hline
         GN20 & 2.62 & $1.78\times1.65$ & -92.5 & 0.32 & 2.5\\
         \\
         ID141 & 2.23 & $1.84\times1.59$&-101  & 0.40 & 8.6\\
         \\
         HDF850.1 & 2.45 & $2.13\times1.98$&73.3 & 0.36 & 0.2\\
         \\
         PSSJ2322+1944 & 5.17 & $1.26\times 1.24$&18.6 & 0.44 & 2.6\\
         \\
         J2054-0005 & 3.45 & $1.57\times0.79$ & 197 & 0.43 & 0.1\\
         \\
         J2310+1855 & 3.34 & $1.32\times0.95$ & 194 & 0.34 & 0.8\\
         \hline
    \end{tabular}
    \tablefoot{(1) Source name. (2) Integration time (6-antennas equivalent). (3)  Beam (major $\times$ minor axis). (4) Beam position angle.  {\color{black}(5) Achieved line sensitivity. (6) Achieved continuum sensitivity.}}
    \label{tab:2}
\end{table*}

\section{Sample, observations and data reduction}
\subsection{The sample}

The selected sources for this study consist of three submillimeter galaxies and three quasar host galaxies at 4$\lesssim$z$\lesssim$6 that were observed using the IRAM interferometer. We give a brief description of our sources below. \\

GN20 is one of the brightest and most well-studied SMGs identified in the Great Observatories Origins Deep Survey (GOODS) Northern field (Pope et al. 2006) at the redshift z = 4.055, and lies in a massive proto-cluster environment with, most notably, two companion galaxies, GN20.2a and GN20.2b, and a Lyman Break Galaxy (LBG), dubbed BD29079, which are all separated within a radius of 25$''$ (corresponding to a projected separation of $\sim$180 kpc at z=4.05) centered at GN20 (Daddi et al. 2009). CO(2-1) maps reveal a clumpy, extended gas disk with a diameter of 14 $\pm$ 4 kpc \citep{Hodge2012}. GN20 has a high infrared luminosity and displays signatures of an extreme starburst with a large specific star formation rate (sSFR) excess ($\sim$6x) compared to typical galaxies at the same epoch \citep{Tan2014, Colina2023}. {\color{black}The source, however, has not been observed for the [CII] 158 $\mu$m emission line. \footnote[1]{\color{black}The [CII] emission line for GN20 corresponds to an observed frequency ($\nu_{obs}$) of  $\sim$377 GHz. This frequency coincides with a significant atmospheric feature centered around 380 GHz, which precludes coverage by the IRAM receiver. Additionally, the source’s declination of +62$^{\circ}$ places it outside the observable range of ALMA.}}\\



ID141 is a highly luminous lensed source observed in the Herschel Astrophysical Terahertz Large Area Survey (H-ATLAS) project (Eales et al. 2010). It has an extreme intrinsic brightness and a high intrinsic SFR of $\sim$2000 M$_{\odot}$ yr$^{-1}$. Being one of the brightest SMGs in the sky, it has been subjected to numerous sub-mm scans (\citealt{Cox2011, Dye2022} and references therein). Additionally, high spatial resolution ALMA observations of this source mapping CO(7-6) \& H$_2$O(2$_{1,1}$ -- 2$_{0,2}$) emission lines reveal kinematics consistent with a rotating disk system and display evidence of a nearby perturbed component, suggesting the source is either a rotating disk galaxy with a prominent outer ring or a merging system with two or more close components \citep{Dye2022}. \\

HDF850.1 is the brightest SMG in the confusion-limited James Clerk Maxwell Telescope (JCMT)/Submillimetre Common-User Bolometer Array (SCUBA) survey of the northern Hubble Deep Field \citep{Hughes1998}. Its redshift, z = 5.183, was measured from the CO(5-4) and (6-5) transitions and confirmed by the detection of the [CII] 158 $\mu$m emission line, all observed using the Plateau de Bure Interferometer (Walter et al. 2012). Inspection of the redshift distribution of galaxies around HDF850.1 reveals that HDF850.1 lies in an overdense environment \citep{Walter2012}, and higher resolution [CII] 158 $\mu$m observations reveal a galaxy merger at play \citep{Neri2014}. {\color{black}Furthermore, recent JWST NIRCam observations have detected an H$\alpha$ emission line at a very high significance (Herard-Demanche et al. 2023) and resolved the UV–optical counterpart \citep{Sun2024} from this highly obscured source.} 
\\

PSSJ2322+1944 is a strongly lensed hyperluminous quasar at z = 4.12 that appears as an Einstein ring in CO(2-1) maps \citep{Carilli2003}. Identified in a spectroscopic follow-up of the Palomar Sky Survey \citep{Djorgovski2000}, it is one of the brightest known sources at z$\sim$4 both in CO and infrared luminosity and has been detected in multiple CO transitions \citep{Cox2002,Carilli2002}; however, only upper limits were reached on the [CII] 158 $\mu$m line \citep{Pety2004}.\\

J2054-0005 and J2310+1855 are z$\sim$6 quasars first discovered from the Sloan Digital Sky Survey \citep[SDSS;][]{Jiang2009, Jiang2016}. With a black hole mass of 0.9 $\times$ 10$^9$ M$_{\odot}$ (2.3 $\times$ 10$^9$ M$_{\odot}$) and absolute magnitude M$_{1450}$ of -26.1 (27.8) for J2054-0005 (J2310+1855) respectively, they are some of the most extensively studied quasars at cosmic dawn and were detected in multiple sub-mm transitions, including [CII] 158 $\mu$m, [CI] 369 $\mu$m, [OIII] 88 $\mu$m, OH 119, 163 $\mu$m, and multiple CO and H$_2$O lines \citep{Wang2013, Decarli2018, Hashimoto2019, Li2020, Pensabene2021, Tripodi2022}. 

\subsection{Observations and data reduction }
The observations were performed during the winter semesters of 2011, 2012, and 2016 using the IRAM interferometer with six (and seven) antennas in the compact C or D configurations. The quasars J2054-0005 and J2310+1855 were observed as part of the project W16EF (PIs: Ferkinhoff \& Decarli) with NOEMA, while the remaining four sources, {\color{black}included in the projects V0B2 and W0B7, were observed with PdBI} (PI: Decarli). The WideX Band 3 \& 4 receivers were used for all the observations, covering the 1 and 0.8 mm atmospheric windows to detect the redshifted [NII] 205 $\mu$m emission line and the underlying dust continuum emission. All data were processed with the \textsf{clic} software within the \textsf{GILDAS} suite and imaged with the \textsf{mapping} package. \\




The QSO PSSJ2322+1944 was observed on {\color{black}2011 December 26 and 28, 2012 December 12, and 2013 May 14, 26, and 27}. The blazar 3C454.3 was used for phase and amplitude calibration, while the radio star MWC349 was used for absolute flux calibration. The first track was observed under
unfavorable weather conditions (precipitable water vapor, pwv = 2-4 mm,
system temperature, $T_{\rm sys}$ = 350-450 K); the second and third
tracks were secured under much better weather (pwv=0.7-2.0 mm, $T_{\rm
sys}$ = 160-250 K). We reached 5.17 hrs of integration on-source time. \\

The SMG HDF850.1 was observed on {\color{black}2012 January 10 and 11}. The radio-loud source 1150+497 was observed as a phase and amplitude calibrator, while 3C279 and MWC349 were observed for pointing and flux calibration. The weather conditions were good (pwv=1-2, $T_{\rm sys}$=150-250 K on January 10; pwv=2-3 mm, $T_{\rm sys}$=160-230 K). We reached an integration time of 2.45 hr on source. \\


\begin{figure*}
    \centering
    \includegraphics[width=18.5cm]{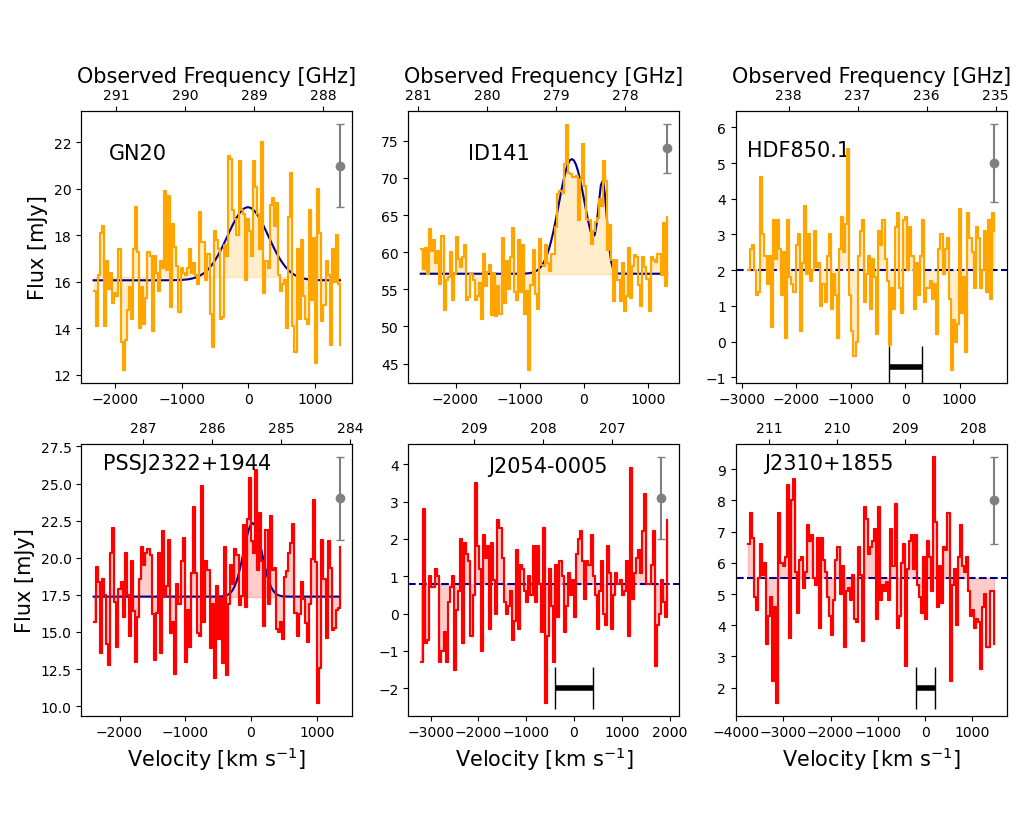}
    \caption{Observed [NII] 205 $\mu$m spectra of the sources in our sample. The orange histogram represents the 1-D spectra of submillimeter galaxies, while the red histogram represents the spectra of quasar hosts. The best Gaussian + continuum fit is shown in solid blue. {\color{black}The dashed blue line represents the mean dust continuum flux for sources undetected in the [NII] 205 $\mu$m emission}. {\color{black}The gray bar represents the mean error (RMS) of the spectrum, while the black bar represents the width of the [CII] 158 $\mu$m line used to estimate the upper limit on the integrated line flux for the non-detections (see Sect. 3.1).} The zero velocity is with respect to the reference redshift in Tab. 1.}
    \label{fig:spectra}
\end{figure*}

The SMG GN20 was observed on {\color{black}2012 January 15}, using 1044+719 and 1300+580 as phase and amplitude calibrators, and 3C345 and 2200+420 as flux and pointing calibrators. The weather conditions were good
(pwv=1.0-1.8 mm, $T_{\rm sys}$ = 200-250 K). These observations resulted in an on-source time of 2.62 hrs.\\

The SMG ID141 was observed on {\color{black}2012 January 16}. The calibrators 1502+106
and 1402+044 served as phase and amplitude reference, and various
sources, including MWC349, 1502+036, and 1055+018, were used to set the
flux scale. The weather conditions were nominal (pwv = 1.5-2.0 mm,
$T_{\rm sys}$ = 180-250 K). The integration time reached on the source is 2.23 hrs.\\

The two z$\sim$6 QSOs, J2054-0005 and J2310+1855, were observed on {\color{black}2016 December 1 and 3} and {\color{black}2016 November 17, 19, and 26}, respectively, when the array had seven antennas. The blazar 3C454.3 served as a bandpass, amplitude, and phase calibrator for both sources, while MWC349 was used for the absolute flux calibration. System temperature was $T_{\rm sys}$ = 150-250 K throughout the observations. The on-source time for J2054-0005
and J2310+1855 are 3.34 and 3.45 hours (equivalent 6-antenna integration time), respectively. 
\\




The cubes were generated using 30 MHz channels, resulting in velocity channels with widths varying with frequency. The corresponding integration time, spatial resolution (obtained using natural visibility weights), {\color{black}and the observational sensitivities achieved} for each source are mentioned in Tab. \ref{tab:2}.



\section{Results and discussion}
In the following section, we present the results of the [NII] 205 $\mu$m observations and estimate {\color{black}two key ratios} for our sample, {\color{black}namely the [NII]-to-IR \& [CII]-to-[NII] ratios}. We summarize the [NII] 205 $\mu$m emission line properties obtained for each source in Section 3.1. In Sections 3.2 and 3.3, we estimate the [NII]-to-IR as well as the [CII]-to-[NII] line ratio for our sources, respectively, and place them into context with other sources detected in the [NII] 205 $\mu$m emission both at low and high redshifts.

\begin{table*}[!htp]
    \caption{[NII] 205 $\mu$m line properties and luminosity ratios. }
    \label{table: 3}
    \centering
    \begin{tabular}{c c c c c c c c c c}
    \hline
    Source  & $F_{cont}$ & $F_{line}$ & FWHM & z$_{[\mathrm{NII}]}$ & L$_{[NII]}$  & L$_{[NII]}$/L$_{IR}$ & L$_{[CII]}$/L$_{[NII]}$& $f($[CII]$^{PDR})$ \\
           & [mJy] & [Jy km s$^{-1}$] & [km\,s$^{-1}$] && [$10^8$L$_{\odot}$] & [$10^{-5}$]&&&\\
    (1) & (2) & (3) & (4) & (5) & (6) & (7) & (8)  &(9) \\
    
    \hline
    
    GN20  & 16.2 $\pm$ 2.5 & 2.3 $\pm$ 0.7 & 710 $\pm$ 170& 4.0541 $\pm$ 0.0010  & 9.2 $\pm$ 2.8 & 6.1 $\pm$ 1.9 & ... & ... \\
    \\

    ID141  & 57.0  $\pm$ 8.6 & 9.4 $\pm$ 2.0 & {810 $\pm$ 40}& 4.2445 $\pm$ 0.0030& 6.9 $\pm$ 1.6 & 4.4 $\pm$ 1.1 & 15$\pm$4 & 81\%$\pm$5\% (42\%$\pm$16\%)\\
    \\
    
    HDF850.1  & 2.0 $\pm$ 0.2 & < 0.38$^{\dag}$ & 300$^{\ddag}$ & ... &< 0.88$^{\dag}$ & <4.4 & >12.7 &  >76\% (>27\%) \\
    \\
    
    PSSJ2322  & 17.4 $\pm$ 2.6 & 1.7 $\pm$ 0.7 & 330 $\pm$ 110& 4.1190 $\pm$ 0.0007 & 1.3 $\pm$ 0.6 & 2.4 $\pm$ 1.2 & <  14.1 & <78\% (<35\%)\\
    \\

    J2054-0005 & 0.8 $\pm$ 0.1 & < 0.3$^{\dag}$  &  200$^{\ddag}$& ... & < 2.19$^{\dag}$ &<2.4 & >15.3    & >80\% (>41\%)  \\
    \\

    J2310+1855  & 5.5 $\pm$ 0.8 &  < 0.5$^{\dag}$ &  400$^{\ddag}$& ... & <  3.47$^{\dag}$ &<1.5 & >23.9  & >87\% (>62\%)        \\
    \hline
    
    \end{tabular}
    \tablefoot{(1) Source name. (2) Continuum flux density at the frequency of the redshifted reference frequency (Tab. 1). (3) Integrated line flux. (4) FWHM of the [NII] 205 $\mu$m line. (5) Measured [NII] redshift. (6) [NII] 205 $\mu$m luminosity. (7) [NII]-to-IR luminosity ratio. (8) [CII]-to-[NII] luminosity ratio. (9) Fraction of [CII] 158 $\mu$m emission arising from photon-dominated regions (PDRs) assuming [CII]$^{ion}_{158 \mu m}$/[NII]$_{205 \mu m}$ $\sim$3 ($\sim$9; see Sect. 3.3). The errors on fluxes and integrated lines were scaled up by 15\% to account for possible calibration uncertainties. {\color{black}The flux densities are apparent values, while all the luminosities are corrected for gravitational magnification.}\\
    $^{\dag}$3$\sigma$ upper limit on the integrated flux and luminosity. \\
    $^{\ddag}$ [CII] 158 $\mu$m FWHM used to calculate the limits on the integrated flux and luminosity.}
\end{table*}


\subsection{[NII] 205 $\mu m$ line emission}
Most of the sources are spatially unresolved in our observations,
with the exception of ID141, but their continuum was detected with high significance. {\color{black}Fig. 1 shows the spectra extracted from an aperture enclosing the 1$\sigma$-region in the [NII] 205 $\mu$m integrated line maps.  For non-detections, the 1$\sigma$-region is taken from the continuum maps. We simultaneously fitted the emission lines with a Gaussian profile and the underlying continuum and converted the integrated fluxes into line luminosity using the following equation}:

\begin{equation}
    L_{[NII]} = 1.04 \times 10^{-3} \times F_{line} D_L^2 \nu_{obs} \hspace{5px} \mathrm{L}_{\odot}
\end{equation}
Where {\color{black}$F_{line}$} is the integrated flux of the line in Jy km s$^{-1}$, $D_L$ is the luminosity distance in Mpc, and $\nu_{obs}$ is the observed frequency of the line in GHz (see, e.g., Carilli \& Walter 2013). The luminosities and derived fit parameters are listed in Tab. \ref{table: 3}.\\ 

{\centering GN20 \\}
The [NII] 205 $\mu$m emission line is clearly detected in this SMG, displaying a single Gaussian line with a line flux of $F_{line}$ = 2.3 $\pm$ 0.7 km s$^{-1}$ and a large line width of 710 $\pm$ 170 km s$^{-1}$. We detected the underlying continuum with a flux density of {16.2 $\pm$ 2.5} mJy. The line width is comparable to the width measured from CO emission lines \citep{Carilli2010, Carilli2011, Hodge2012}.\\ 

{\centering ID141\\}

The [NII] 205 $\mu$m emission line is also well detected for ID141, displaying a clear double peak profile. We, therefore, fit the spectrum with a double Gaussian to obtain the line parameters. We estimated an integrated flux of 7.6 $\pm$ 0.8 Jy km s$^{-1}$ (FWHM = 470 $\pm$ 60 km s$^{-1}$) and 1.8 $\pm$ 0.4 Jy km s$^{-1}$ (FWHM = 160 $\pm$ 40 km s$^{-1}$) from the fit and a continuum flux density of 57.0 $\pm$ 0.8 mJy. Our measurement of the total integrated flux, $F_{line}$ = 9.4 $\pm$ 2.0 Jy km s$^{-1}$, is consistent with the value derived by Cheng et al. (2020) ($F_{line}$ =  7.9 $\pm$ 1.9 Jy km s$^{-1}$) observed using ALMA, while the FWHM of the total [NII] line emission is consistent, within the uncertainties, with the FWHM measurements from Cheng et al. (2020) as well as [CII] 158 $\mu$m and CO lines from Cox et al. (2011).\\



{\centering PSSJ2322+1944 \\}
The [NII] 205 $\mu$m emission line is well detected, displaying a single Gaussian line with an integrated line flux of $F_{line}$ = 1.7 $\pm$ 0.7 Jy km s$^{-1}$ and a FWHM of 330 $\pm$ 110 km s$^{-1}$, which aligns well with the widths observed in CO emission lines (Cox et al. 2002). The underlying continuum is detected with a flux density of {17.4 $\pm$ 2.6} mJy.\\

{\centering Non-detections \\}
The [NII] $205\, \mu $m emission is undetected for HDF850.1, J2054-0005, and J2310+1855 from our observations. To provide an upper limit to the integrated line flux of the sources with undetected [NII] $205\, \mu$m emission, we integrated the spectrum within the FWHM of the [CII] $158\, \mu $m line for the same redshift. The continuum flux density, upper limits on the integrated flux, and the corresponding [NII] 205 $\mu$m luminosities are mentioned in Tab. \ref{table: 3}.


\subsection{Line deficit} 
\begin{figure}
    \centering
    \resizebox{\hsize}{!}{\includegraphics{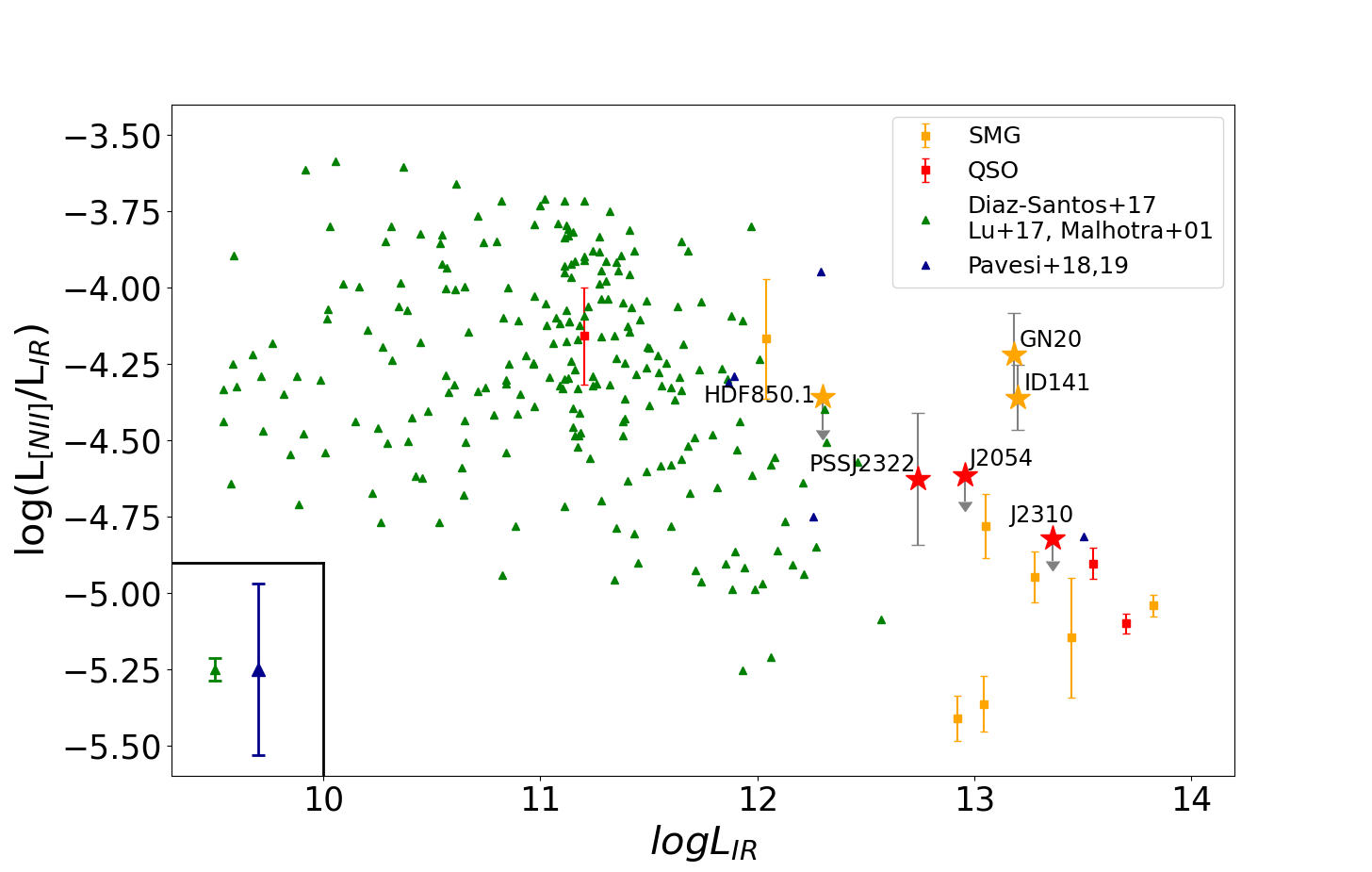}}
    \caption{The [NII]-to-IR ratio as a function of IR luminosity for the sources in our sample. Our sample is illustrated as stars, following the same color scheme in Fig. 1. We also compile sources from the local Universe (green triangles) as well as at high redshifts (orange/red squares \& blue triangles) that have been detected in [NII] 205 $\mu$m emission. The green and blue bars represent the mean error bars for their respective data points. All luminosities are corrected for gravitational magnification. Our new measurements further populate the sparser high-end of the infrared luminosity regime of such sources. }
    \label{fig:2}
\end{figure}

In Fig. \ref{fig:2}, we compare the [NII] and IR luminosities of our sources. {\color{black}We computed the infrared luminosities by integrating the dust emission model across the 8-1000 $\mu$m wavelength range (see Appendix B for more details on the modeling of the dust spectral energy distribution)}. We plot the [NII]/IR values against the IR luminosity for our sources and compare them with other SMGs and QSOs detected in the [NII] 205 $\mu$m at 3 < z < 6 \citep{Decarli2014, Bethermin2016, Pavesi2016, Lu2017_b, Lu2018, Tadaki2019, Doherty2020, Schreiber2021} as well as with galaxies in the local Universe \citep{Malhotra2001, Diaz-Santos2017, Lu2017_a}. We also include a set of normal galaxies between z=5-6 taken from \cite{Pavesi2018,Pavesi2019}. We converted the [NII] 122 $\mu$m values from Malhotra et al. (2001) and D{\'\i}az-Santos et al. (2017) into [NII] 205 luminosities using a L$_{[NII]_{122}}$/L$_{[NII]_{205}}$ ratio of $\sim$ 5 \citep[see, e.g.,][]{Beirao2010, Decarli2012}. \par

We observed a decrease in [NII]-to-IR ratio by an order of magnitude, both in local galaxies and high-z sources, as we move towards the brighter end of the dust continuum luminosity. This trend has been referred to as the [NII] fine structure line deficit \citep{Gracia-Carpio2011, Farrah2013, Diaz-Santos2017, Zhao2013, Zhao2016}.  While the exact causes of the deficit are still under debate, we explored a few possible physical mechanisms responsible for the observed trend. \par 
1) High ionization parameter: Using the radiative transfer code CLOUDY \citep{Ferland1998} and an isobaric, planar slab geometry, \cite{Abel2009} showed that a higher ionization parameter (ionizing radiation to particle density ratio) leads to a larger fraction of ionizing photons absorbed by dust within the HII region, and thereby to less heating and line cooling {\color{black}(We recommend that the reader refer to Section 1. in Abel et al. (2009) for a physical explanation of this phenomenon)}. A larger ionization parameter could, therefore, explain the smaller line-to-continuum flux ratio observed in local ULIRGs compared to less luminous IR-bright galaxies. Similarly, \cite{Gracia-Carpio2011} observed a drop in the [CII] 158 $\mu$m as well as other fine-structure lines to FIR continuum ratios (see Fig. 2 of Graci{\'a}-Carpio et al. 2011) by a factor of 10 for galaxies with high L$_{FIR}$/M$_{H_2}$ ratio in a sample of 44 sources ranging from local starbursts, Seyfert galaxies, and infrared luminous galaxies both at low and intermediate redshifts in the {Herschel}-PACS {SHINING} survey\footnote[2]{The Survey with Herschel of the ISM in Nearby INfrared Galaxies {(SHINING)} survey was a comprehensive far-infrared spectroscopic and photometric survey of bright infrared galaxies and AGNs at local and intermediate redshifts, conducted using the {Herschel} Photodetector Array Camera and Spectrometer (PACS) \& Spectral and Photometric Imaging Receiver (SPIRE) instruments.}. This drop seems to be a universal feature of galaxies irrespective of their redshift and optical activity. To explain this deficit, Graci{\'a}-Carpio et al. extended Abel's analysis to include the other fine structure lines observed in the PACS survey. From the model, they were able to reproduce the deficit in the [CII] 158 $\mu$m, [NII] 122 $\mu$m, and [OI] 63, 145 $\mu$m lines for galaxies with L$_{FIR}$/M$_{H_2} > 80$  L$_{\odot}$/M$_{\odot}$. Since to first order, the ionization parameter is proportional to L$_{FIR}$/M$_{H_2}$, increasing the ionization parameter by at least an order of magnitude from the typical value (U$\sim$$10^{-3}$), which explains the average line to FIR ratios measured in galaxies with low L$_{FIR}$/M$_{H_2}$, the model was able to reproduce the deficits consistent with the observed trend. With L$_{FIR}$/M$_{H_2}$ ratios > 80 L$_{\odot}$/M$_{\odot}$ for our sources (see Tab. 1), a higher ionization parameter could be one of the main contributors to the observed deficit.\par



2) Thermalization: The [CII] deficit is found to be stronger in the ionized medium than compared to the neutral medium \citep{Sutter2019}. \cite{Sutter2021} observed that in the ionized ISM, the [CII]/IR ratio plummets when the electron number density is close to the [CII] critical density, indicating that thermalization could be a contributing factor to the line deficit. This scenario may also be extended to the [NII] 205 $\mu$m deficit since it arises solely in the ionized phase of the ISM. While thermalization is very likely to contribute in high density environments since the critical density of [NII] 205 $\mu$m is $\sim$50 cm$^{-3}$, however, the exact extent to which it plays a role to the observed [NII] 205 $\mu$m line deficit can only be explored once a significant sample of high-z galaxies observed both in the [NII]  205 $\mu$m emission line and constraints on the electron number density\footnote[3]{The electron density of high-z galaxies can be estimated using optical tracers such as [OII] (3729\AA/3727\AA) and [SII] (6718\AA/6732\AA) or with the [NII] 122 $\mu$m line in combination with the [NII]  205 $\mu$m emission line. }  of the ISM become available.\par 
3) Dusty HII regions: Another possible explanation is a higher abundance of dust in HII regions found in IR-bright galaxies \citep{Farrah2013, Diaz-Santos2017, Herrera-Camus2018}. In this case, a higher fraction of UV photons produced by young massive stars is absorbed by dust, which increases the IR emission and thus suppresses the fraction of ionizing photons that photoionize hydrogen in the HII regions, thereby decreasing the line emission relative to the IR luminosity. \par 

 While the combination of the aforementioned physical mechanisms may be responsible for the observed deficit, it should be noted that such a line deficit is expected, as explained in Walter et al. (2022), since the infrared luminosity scales proportional to the dust temperature with a power-law exceeding the Stefan–Boltzmann law (L$_{dust}$$\sim$T$^{4.6}$). In contrast, the line emission typically scales linearly (L$_{line}$$\sim$T), and no known excitation mechanism could allow the line flux to keep up with the strong temperature dependence of the dust luminosity. Furthermore, when we split the sources in our sample and those detected between z=3-6 into QSOs and SMGs, we find that the observed [NII]-to-IR ratios are indistinguishable, suggesting that the AGN has little impact on the [NII] 205 $\mu$m emission in these sources.

\begin{figure}
    \centering
    \resizebox{\hsize}{!}{\includegraphics{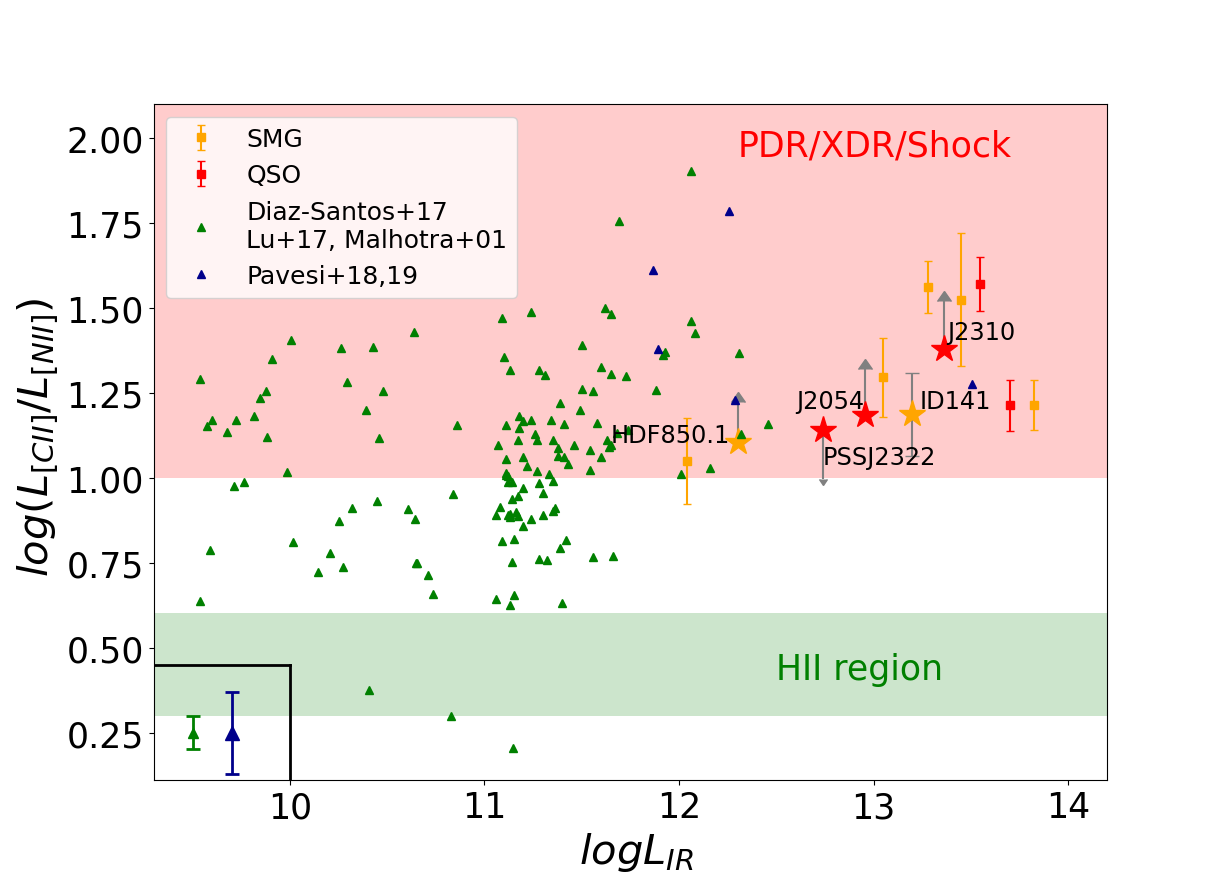}}
    \caption{The [CII]-to-[NII] ratio as a function of IR luminosity for the sources in our sample as well as for galaxies taken from the literature that are detected in both [CII] 158 $\mu$m and [NII] 205 $\mu$m, assuming the {\color{black}commonly adopted [CII]$^{ion}_{158 \mu m}$/[NII]$_{205 \mu m}$ value of $\sim$3, following the study from \cite{Oberst2006}.} The color scheme adopted is the same as in Fig. 2. Most of the sources in our sample show a value $\geq$10, indicating that the bulk of the [CII] emission of our sources arises from the neutral medium.}
    \label{fig:3}
\end{figure}

\subsection{The [CII]-to-[NII] ratio }

The [CII]-to-[NII] ratio is a powerful diagnostic that provides information on the metal enrichment of the ISM and probes the abundance of the C$^+$ in the ionized medium. Studies have used the [CII]-to-[NII] line ratio to investigate the different properties of the ISM in high-z sources, ranging from Ly$\alpha$ emitters (LAEs) to SMGs and QSOs \citep{Decarli2014, Lu2017_b, Cunningham2020, Pensabene2021}. The estimates on this line ratio expose the diversity in global properties and conditions of the ISM in various galaxies. In Fig. \ref{fig:3}, we plot the [CII]-to-[NII] ratio as a function of IR luminosity for our sources, together with SMGs and QSOs at 3<z<6 that have been detected both in [CII] 158 $\mu$m and [NII] 205 $\mu$m \citep{Wagg2010, Pavesi2016, Bethermin2016, Lu2017_b, Schreiber2018, Tadaki2019} and local galaxies. Due to the lack of information on the [CII] 158 $\mu$m emission for GN20, we excluded it from the analysis. We find that for all of our sources, the [CII]/[NII] value is >10, corresponding to the XDR{\footnote[4]{XDR: X-ray dominated region}}/PDR and shock models determined by \cite{Decarli2014} and consistent with the measurements of IR–bright sources \citep{Decarli2014, Pavesi2016, Cunningham2020}. For comparison, a [CII]/[NII] value of $\sim$2-4 is expected in HII regions \citep{Decarli2014}. The observed values imply that the neutral medium is the dominant source of [CII] emission for the sources mentioned here. We, therefore, estimate the fraction of the [CII] 158 $\mu$m emission arising from the PDRs by assuming [CII]$^{ion}_{158 \mu m}$/[NII]$_{205 \mu m}$ $\sim$3, a value that is typically observed in HII regions:

\begin{equation}
    f{([\mathrm{CII}]^{PDR})} \approx 1-3 \frac{[\mathrm{NII}]_{205 \mu m}}{[\mathrm{CII}]_{158 \mu m}}
\end{equation}

We found that the [CII]$^{PDR}$ fraction ranges from 76\% to $>$87\% for our sample. However, more recent estimates give a higher [CII]$^{ion}_{158 \mu m}$/[NII]$_{205 \mu m}$ $\sim$9 ratio (Decarli et al. 2023). Adopting this value, the estimated [CII]$^{PDR}$ fractions lower to 27\%-62\% for our sources. The estimated fractions for each source is mentioned in Tab. 3. Similar to the [NII]-to-IR ratios, we find no notable difference in the [CII]/[NII] values between the two galaxy populations, further strengthening that the AGN has a negligible influence on the [NII] 205 $\mu$m emission line. 


\section{Conclusions}

   In this study, we have presented [NII] 205 $\mu$m fine structure line observations in three submillimeter galaxies and three quasar host galaxies at 4$\lesssim$z$\lesssim$6. Three out of the six sources (GN20, ID141, PSSJ2322+1944) are detected in the [NII] 205 $\mu$m emission. While all of our sources are detected in the underlying continuum, for the non-detections in the [NII] 205 $\mu$m emission line, we have placed upper limits on the integrated line flux and, hence, the line luminosity. The [NII]/IR ratios are consistent with other high-z sources and follow the trend of the so-called [NII] 205 $\mu$m line deficit, i.e., the decrease of the line-to-continuum ratio with increasing dust continuum luminosity. While the line deficit is expected to be observed for warmer
sources due to the strong dependence of the dust temperature on the infrared luminosity compared to the line luminosity, other mechanisms such as high ionization parameter, thermalization, and dusty HII regions could also be responsible for the apparent drop in the [NII]/IR ratio. We also find that the [CII]-to-[NII] ratio for our sources lies in the XDR/PDR/Shock regions, consistent with values of other IR-bright sources, indicating that the neutral medium is the dominant source of the [CII] 158 $\mu$m emission for such galaxies. Finally, we find no significant differences in the [NII]-to-IR and the [CII]-to-[NII] ratio estimates between SMGs and QSOs, suggesting that there is a negligible contribution to the boosting of [NII] 205 $\mu$m emission due to AGN photoionization.\par 

   Studies of the [NII] 205 $\mu$m emission have shown it to be a powerful diagnostic for studying the ionized medium of high-redshift galaxies. Future works involving the [NII] 122 $\mu$m, [OIII] 88 $\mu$m fine structure lines and optical diagnostics, observed using JWST, will provide us further insight into a suite of properties of the ionized medium and also reveal the diverse properties between AGN and non-AGN environments.

\begin{acknowledgements}
      We are grateful to the anonymous A\&A referee for their constructive feedback on the manuscript. This work is based on observations carried out under project numbers V0B2 and W0B7, observed with the IRAM Plateau de Bure interferometer, and W16EF, observed with the Northern Extended Millimeter Array. IRAM is supported by INSU/CNRS (France), MPG (Germany), and IGN (Spain). RD acknowledges support from the INAF GO 2022 grant "The birth of the giants: JWST sheds light on the build-up of quasars at cosmic dawn" and from the Italian PRIN 2022 2022935STW - "Black hole formation mechanisms and their impact on high-redshift quasar host properties." This research made use of Astropy (http://www.astropy.org), a community-developed core Python package for Astronomy (Astropy Collaboration 2013, 2018), lmfit (https://lmfit.github.io//lmfit-py/), a python library that provides tools for non-linear least-squares minimization and curve fitting, emcee (Foreman-Mackey, Hogg, Lang, \& Goodman. 2013),  and Matplotlib (Hunter 2007).
\end{acknowledgements}

%
%

\begin{appendix} 
\counterwithin{figure}{section}


\begin{figure*}[h!]

\section{Line integrated and continuum maps}
We present the [NII] 205 $\mu$m integrated emission line and continuum maps of the three submillimeter galaxies and three quasar host galaxies in our sample. The [NII] 205 maps were created by integrating over the line width centering at the redshifted reference frequency, while the continuum maps were created by integrating the cube along the line-free channels. The resulting maps are shown below.

    \centering
    \resizebox{\hsize}{!}{\includegraphics{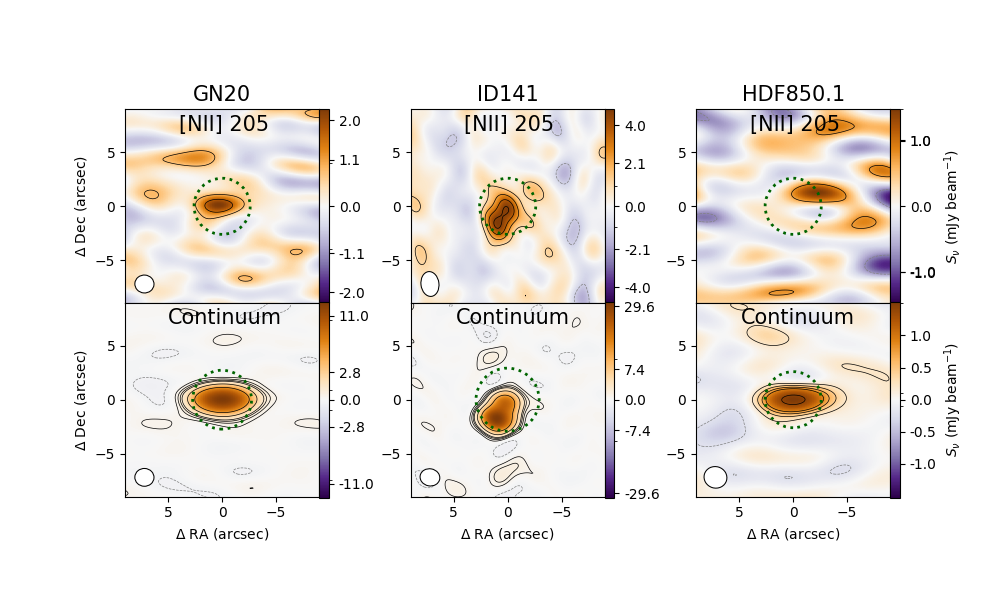}}
    \resizebox{\hsize}{!}{\includegraphics{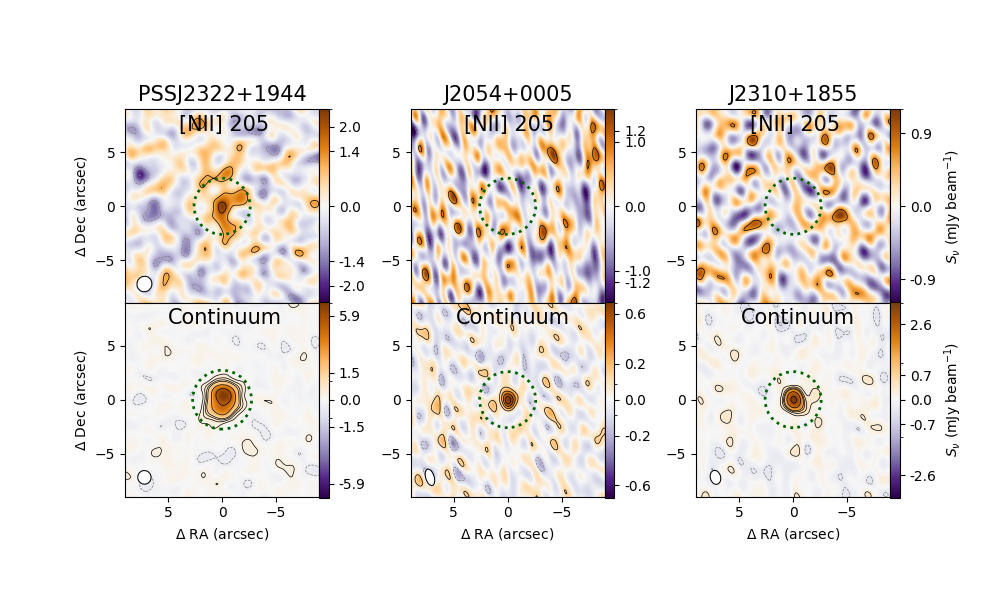}}
    \caption{The [NII] 205 integrated line maps and the corresponding continuum maps of the submillimeter galaxies (upper panel) and the quasar hosts (lower panel) from our sample. The [NII] 205 maps were integrated over the width of their respective [NII] 205 $\mu$m FWHM (and over the [CII] 158 $\mu$m FWHM for sources undetected in the [NII] 205 $\mu$m emission). The full (dashed) contours represent the +(-) 2$\sigma$, 4$\sigma$, 6$\sigma$, 8$\sigma$, 16$\sigma$ regions, where $\sigma$ is the rms noise of the integrated maps. The respective synthesized beam sizes (if present) are shown in the lower left corner of each panel. To guide the eye, a green dotted circle is placed centering the measured continuum coordinates of the source. The respective continuum maps are shown below the [NII] 205 $\mu$m integrated maps.}
    \label{fig:A1}
\end{figure*}

\clearpage

\onecolumn
\section{Dust SEDs}
We reconstruct and fit the dust spectral energy distribution (SED) for our sources, incorporating our new continuum estimates (see Tab. \ref{table: 3}). The continuum flux densities taken from the literature for each source are tabulated in Tab. B.2. We fit the SED by adopting a modified blackbody, following the formula given in Novak et al. (2019) \citep[see][eq. 1]{Novak2019}, and explored the parameter space using a Markov chain Monte Carlo (MCMC) algorithm. Additionally, the sizes of the continuum emitting regions are described in Tripodi et al. (2022) for J2310+1855, and in Wang et al. (2013), Hashimoto et al. (2019), Ishii et al. (2024), and Salak et al. (2024) for J2054-0005. Thus, we model the dust SED of the two quasars in an optically thick regime using a modified blackbody model based on Decarli et al. (2023) \citep[see][eq. 3]{Decarli2023}, where the mean size of the emitting region was fixed during the fits. For both models, we adopt the opacity model following Beelen et al. (2006), which specifies a mass absorption coefficient $k_{0}$ = 0.45 cm$^2$ g$^{-1}$ and $\nu_{0}$ = 250 GHz.\\

The left panels of Fig. B.1. and B.2. display the best fit of the dust SEDs for submillimeter galaxies and quasar host galaxies, respectively. The red star in the plots represents the continuum flux density estimated in this work, while the black squares indicate the flux densities sourced from the literature. On the right of each SED plot, the corner plot shows the posterior probability distributions for the dust mass (M$_{dust}$), dust temperature (T$_{dust}$), and the dust spectral emissivity index ($\beta$). The dashed blue lines in each corner plot mark the 16th, 50th, and 84th percentiles for each parameter, which correspond to a 1-$\sigma$ deviation from the mean. The derived parameters from the fit, along with the estimated total infrared luminosity used in this study, are listed in Tab. B.1.



\begin{table*}[!htp]
\caption{Results of the MCMC SED fitting}
\centering
\begin{tabular}{c c c c c}
\hline
     Source&  M$_{dust}$ & T$_{dust}$ & $\beta$ & L$_{IR}$\\
     & [10$^{9}$M$_{\odot}$] & [K] & & [10$^{13}$L$_{\odot}$]\\
     \hline
     GN20 & 1.99 $\pm$ 0.14 & 32.9 $\pm$ 0.7 & 1.95$^{\ddag}$  & 1.52 $\pm$ 0.11 \\
     \\
     ID141 & 7.67 $\pm$ 0.52 & 38.1 $\pm$ 0.9 & 1.80$^{\ddag}$  & 1.58 $\pm$ 0.12 \\
     \\
     HDF850.1 & 0.23 $\pm$ 0.07 & 30.7 $\pm$ 2.3 & 2.50$^{\ddag}$  & 0.20 $\pm$ 0.05 \\
     \\
     PSSJ2322+1944 & 2.51 $\pm$ 0.37 & 38.1 $\pm$ 3.9 & 1.84 $\pm$ 0.20 & 0.55 $\pm$ 0.10 \\
     \\
     J2054-0005 & 0.17 $\pm$ 0.02 & 59.7 $\pm$ 1.1 & 1.83 $\pm$ 0.07 & 0.90 $\pm$ 0.06 \\
     \\
     J2310+1855 & 0.43 $\pm$ 0.02 & 67.0 $\pm$ 0.3 & 1.88 $\pm$ 0.03 & 2.29 $\pm$ 0.04 \\
     \hline
\end{tabular}
    \tablefoot{(1) Source name. (2) Dust mass. (3) Dust temperature. (4)  Dust spectral emissivity index. (5) Infrared luminosity (8-1000 $\mu$m). \\
    $^{\ddag}$For submillimeter galaxies, the emissivity indices were kept fixed during the MCMC fit and were adopted from Cortzen et al. (2020), Cheng et al. (2019), and Walter et al. (2012) for GN20, ID141, and HDF850, respectively.}
    \label{tab:appendix_1}
\end{table*}

\clearpage


\begin{figure*}
    \centering
    \resizebox{0.93\hsize}{!}{\includegraphics{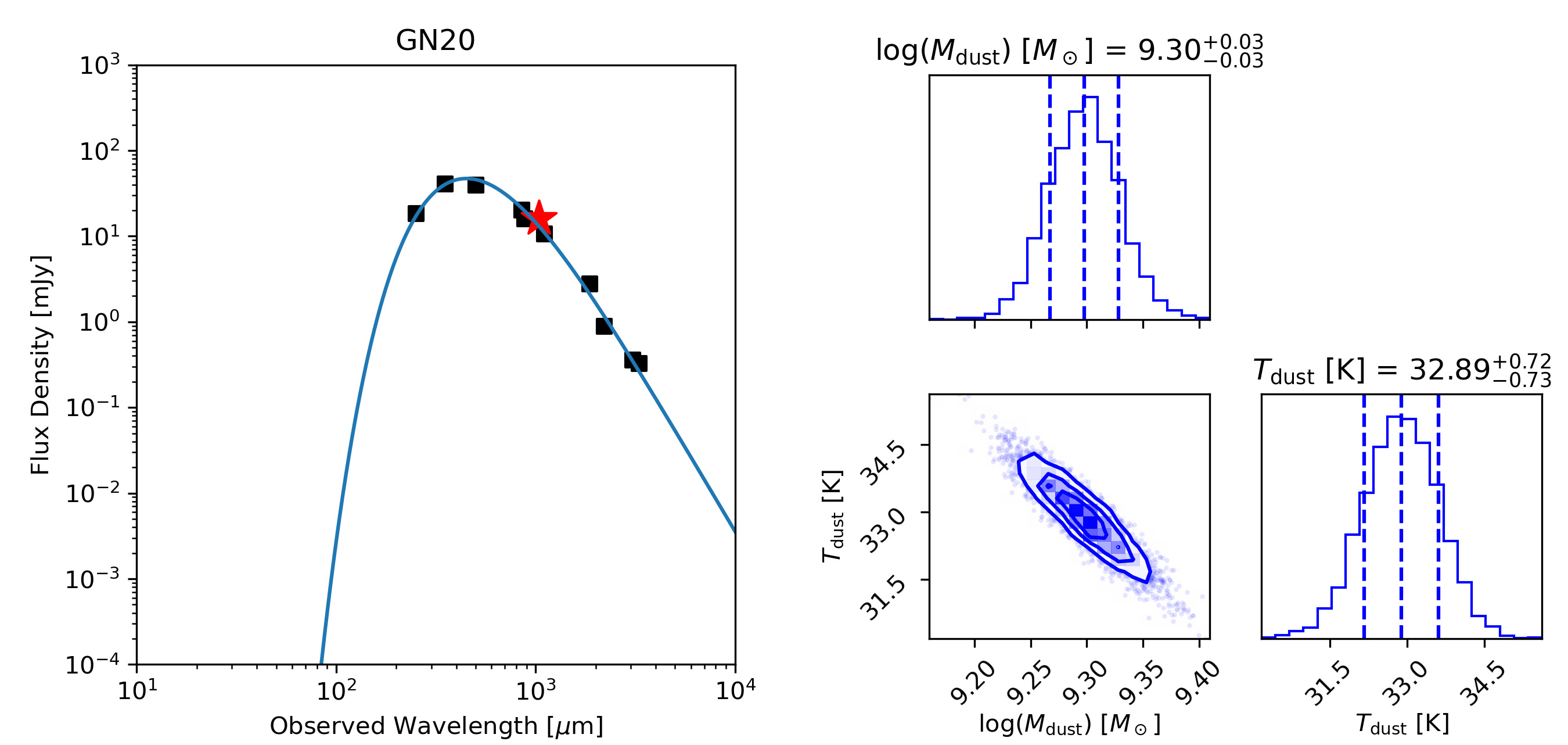}}
    \resizebox{0.93\hsize}{!}{\includegraphics{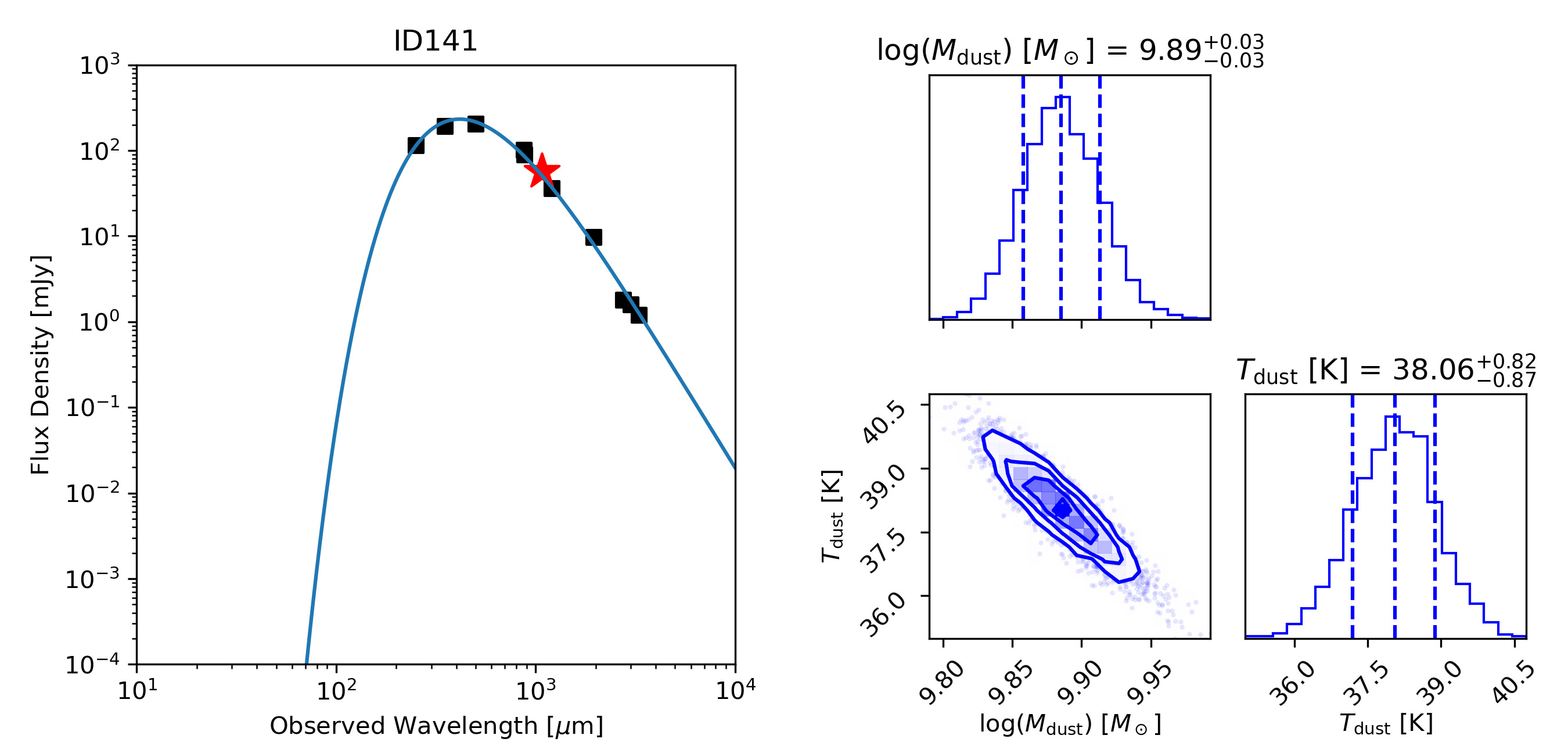}}
    \resizebox{0.93\hsize}{!}{\includegraphics{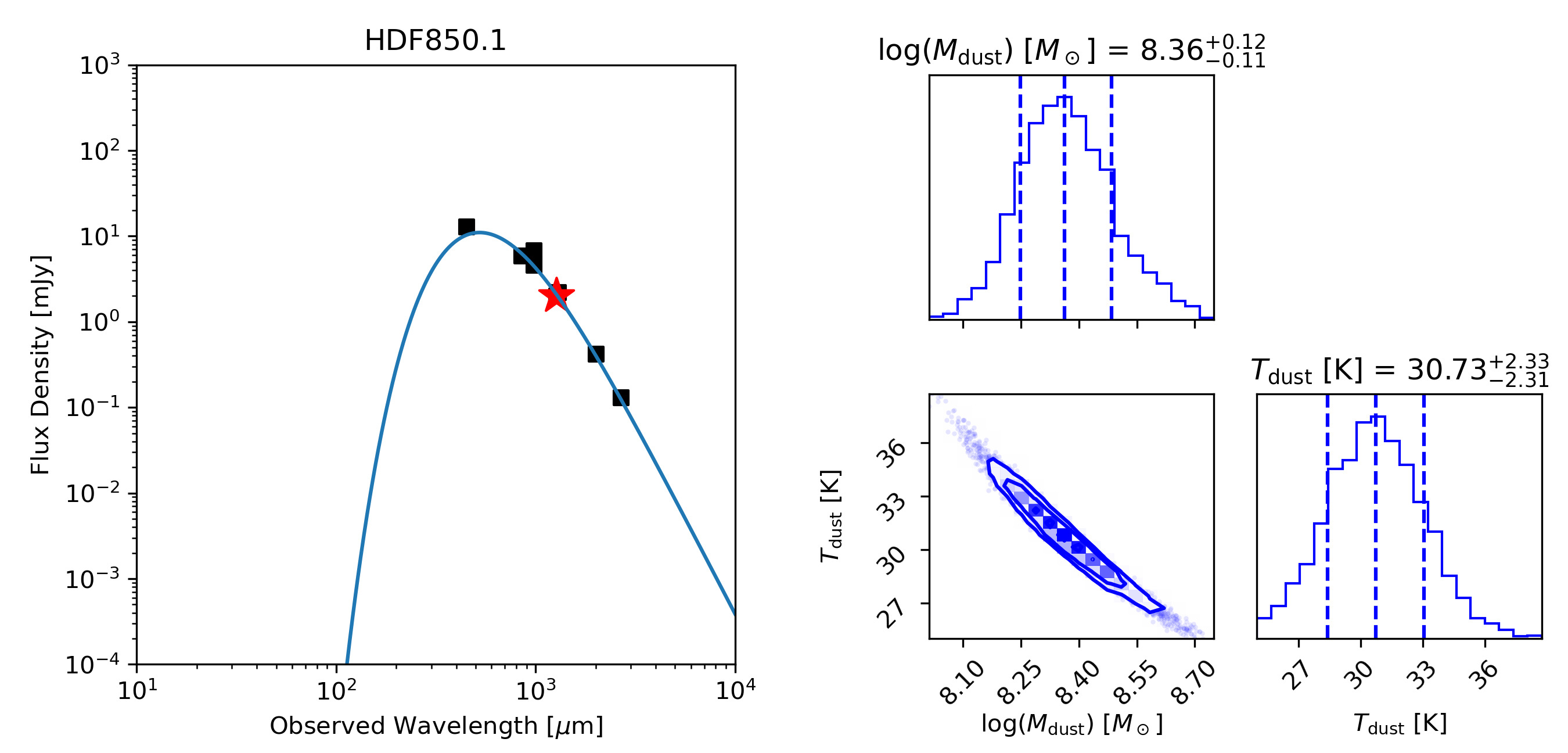}} 
    \caption{Dust SED fit for GN20, ID141, and HDF. See text for more information.}
    \label{fig:appendix_sed_fig_1}
\end{figure*}

\FloatBarrier 

\clearpage

\begin{figure*}
    \centering
    \resizebox{0.95\hsize}{!}{\includegraphics{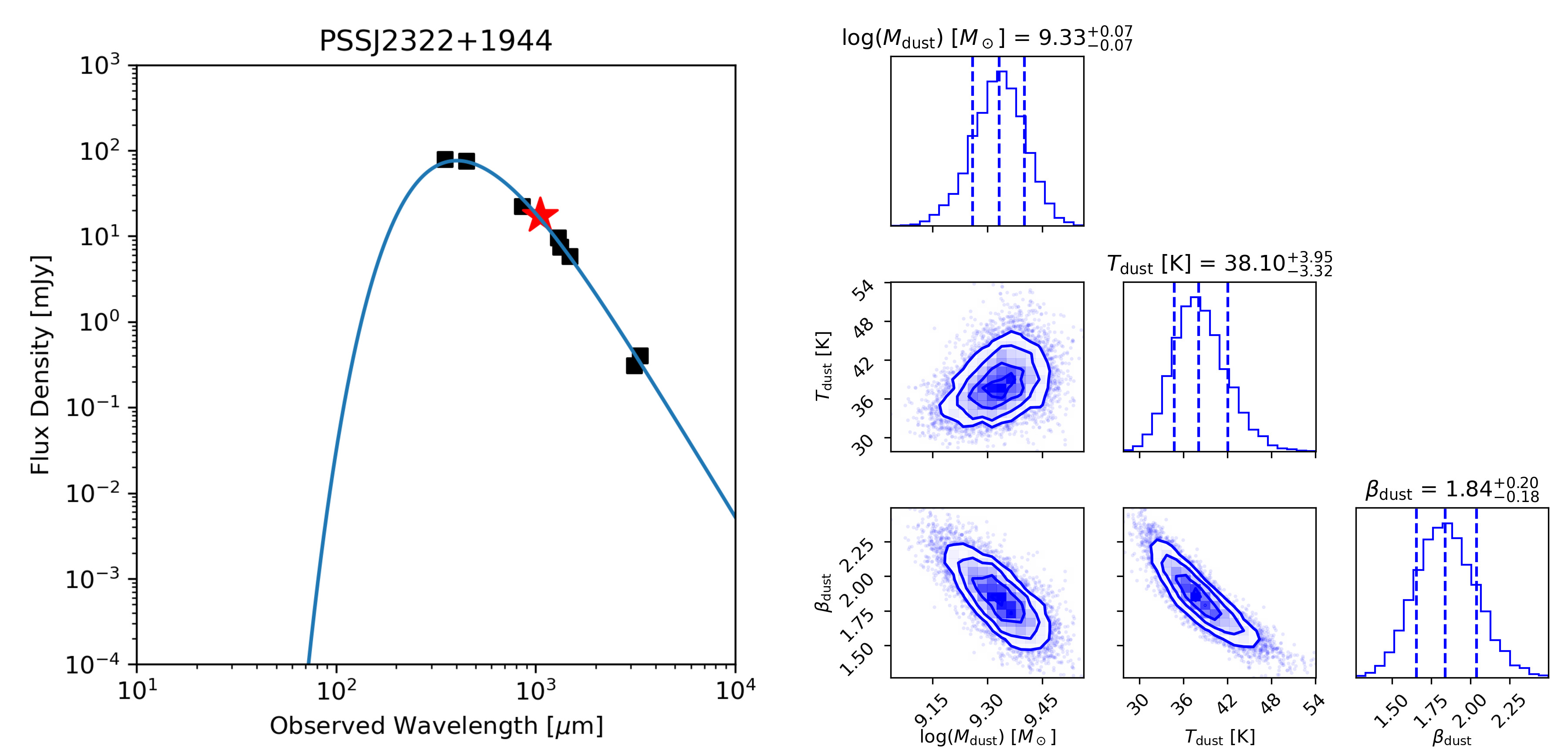}}
    \resizebox{0.95\hsize}{!}{\includegraphics{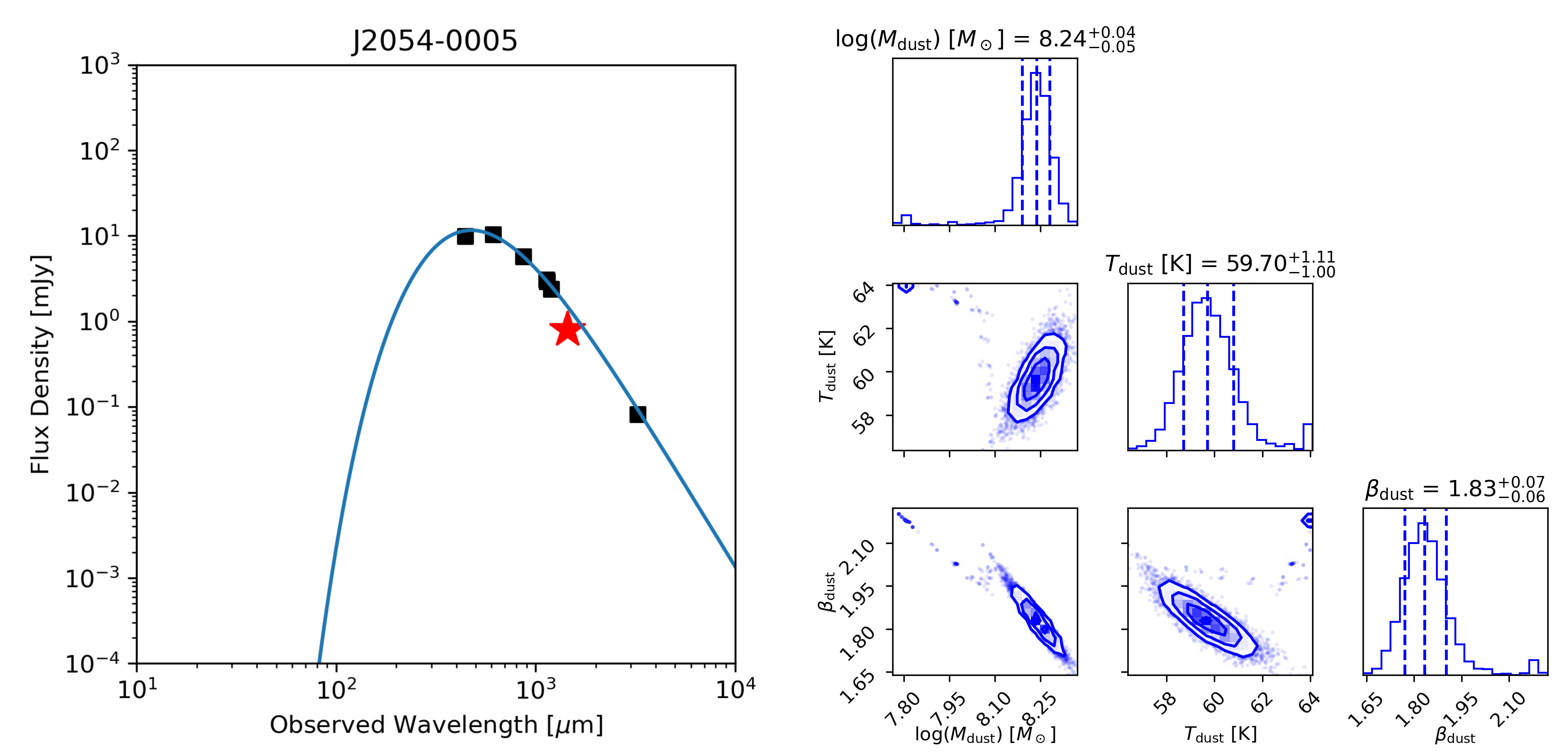}}
    \resizebox{0.95\hsize}{!}{\includegraphics{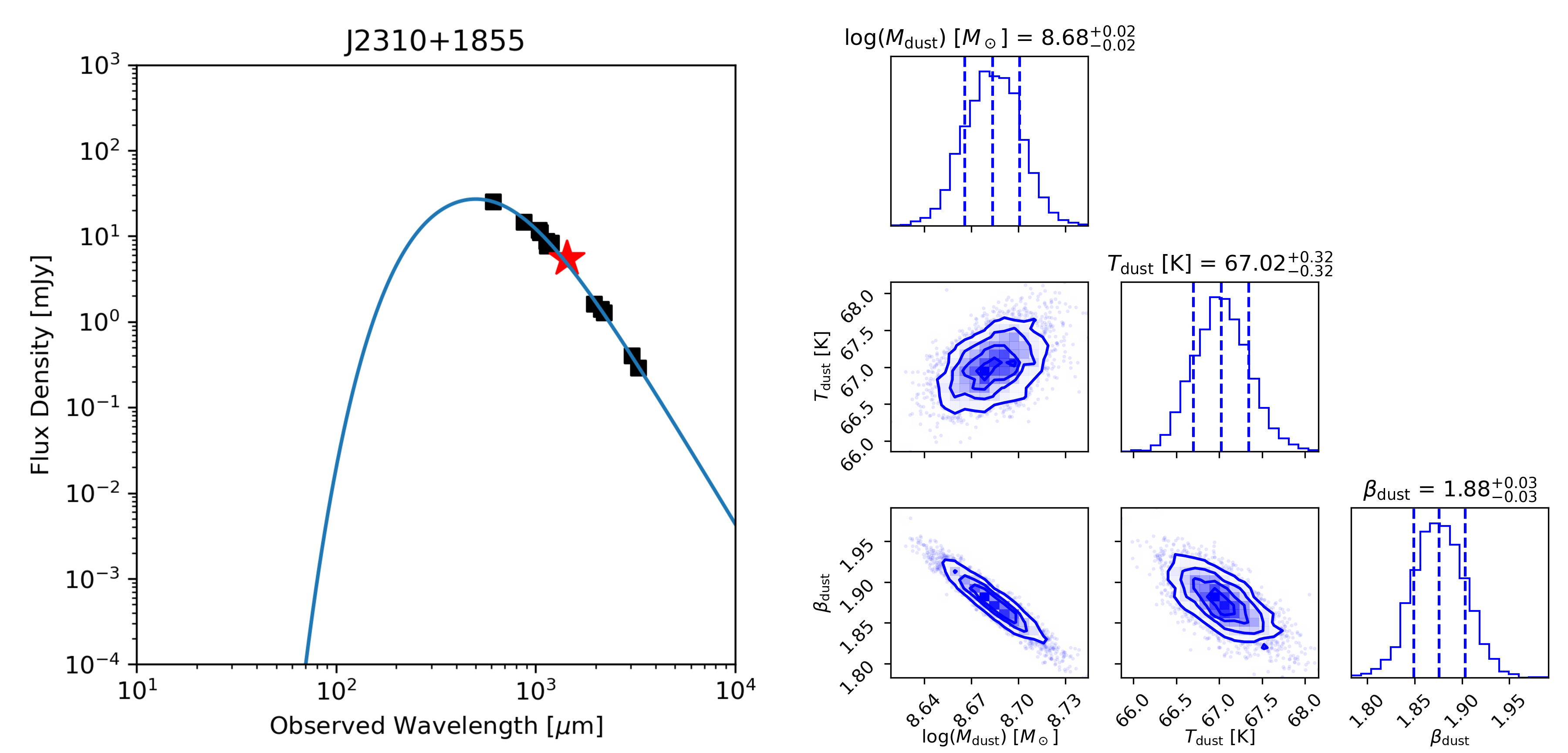}} 
    \caption{Dust SED fit for PSSJ2322+1944, J2054-0005, and J2310+1855. See text for more information.}
    \label{fig:appendix_sed_fig_2}
\end{figure*}

\clearpage

\longtab[2]{
\begin{longtable}{c c c}

    \caption{Continuum flux densities from literature used in the dust SED fit.} \label{table:appendix_2} \\
    \hline
    $\lambda$ & $F_{\lambda}$ & Reference \\
    ($\mu$m) & (mJy) &  \\
    \hline
    \endfirsthead
    \caption{Continued.} \\
    \hline
    $\lambda$ & $F_{\lambda}$ & Reference \\
    ($\mu$m) & (mJy) &  \\
    \hline
    \endhead
    \hline
    \endfoot
    
    \hline

    \multicolumn{3}{c}{{GN20}} \\
    \hline
    250 & 18.6 $\pm$ 2.7 & Magdis et al. 2011 \\
    350 & 41.3 $\pm$ 5.2 & Magdis et al. 2011 \\
    500 & 39.7 $\pm$ 6.1 & Magdis et al. 2011 \\
    850 & 20.3 $\pm$ 2.0 & Pope et al. 2006 \\
    880 & 16.0 $\pm$ 1.0 & Hodge et al. 2015 \\
    1100 & 10.7 $\pm$ 1.0 & Perera et al. 2008\\
    1860 & 2.8 $\pm$ 0.13 & Cortzen et al. 2020\\
    2200 & 0.9 $\pm$ 0.15 & Dannerbauer et al. 2009\\
    3050 & 0.36 $\pm$ 0.05 & Cortzen et al. 2020\\
    3300 & 0.33 $\pm$ 0.06 & Daddi et al. 2009\\
    \hline

    \multicolumn{3}{c}{{ID141}} \\
    \hline
    250 & 115.0 $\pm$ 19.0 & Cox et al. 2011\\
    350 & 192.0 $\pm$ 30.0 & \textquotesingle\textquotesingle  \\
    500 & 204.0 $\pm$ 32.0 & \textquotesingle\textquotesingle  \\
    870 & 102.0 $\pm$ 8.8 & \textquotesingle\textquotesingle  \\
    880 & 90.0 $\pm$ 5.0 & \textquotesingle\textquotesingle  \\
    1200 & 36.0 $\pm$ 2.0 & \textquotesingle\textquotesingle  \\
    1950 & 9.7 $\pm$ 0.9 & \textquotesingle\textquotesingle  \\
    2750 & 1.8 $\pm$ 0.3 & \textquotesingle\textquotesingle  \\
    3000 & 1.6 $\pm$ 0.2 & \textquotesingle\textquotesingle  \\
    3290 & 1.2 $\pm$ 0.1 & \textquotesingle\textquotesingle  \\
    \hline

    \multicolumn{3}{c}{{HDF850.1}} \\
    \hline
    450 & 13.0 $\pm$ 2.7 & Cowie et al. 2017\\
    850 & 5.88 $\pm$ 0.33 & Chapin et al. 2009 \\ 
    975 & 6.8 $\pm$ 0.8 & Walter et al. 2012 \\
    977 & 4.6 & Neri et al. 2014\\
    1300 & 2.2 $\pm$ 0.3 & Downes et al. 1999\\
    2000 & 0.42 $\pm$ 0.13 & Staguhn et al. 2014\\
    2681 & 0.13 $\pm$ 0.3 & Walter et al. 2012 \\
    \hline

    \multicolumn{3}{c}{{PSSJ2322+1944}} \\
    \hline
    350 & 79.0 $\pm$ 11.0 & Beelen et al. 2006 \\
    450 & 75.0 $\pm$ 19.0 & Cox et al. 2002 \\
    857 & 22.5 $\pm$ 2.5 & Carilli et al. 2001 \\
    1298 & 9.6 $\pm$ 0.5 & Omont et al. 2001 \\
    1332 & 7.5 $\pm$ 1.3 & Cox et al. 2002 \\
    1486 & 5.79 $\pm$ 0.77 & Butler et al. 2023 \\
    3123 & 0.31 $\pm$ 0.08 & Pety et al. 2004 \\
    3331 & 0.4 $\pm$ 0.25 & Cox et al. 2002 \\
    \hline

    \multicolumn{3}{c}{{J2054-0005}} \\
    \hline
    444 & 9.87 $\pm$ 0.94 & Tripodi et al. 2024 \\
    612 & 10.35 $\pm$ 0.15 & Hashimoto et al. 2019 \\
    866 & 5.723 $\pm$ 0.009 & Salak et al. 2024 \\
    1136 & 3.08 $\pm$ 0.03 & Tripodi et al. 2024 \\
    1142 & 2.93 $\pm$ 0.07 & Tripodi et al. 2024 \\
    1199 & 2.38 $\pm$ 0.53 & Wang et al. 2008 \\
    3249 & 0.082 $\pm$ 0.009 & Tripodi et al. 2024 \\
    \hline

    \multicolumn{3}{c}{{J2310+1855}} \\
    \hline
    609 & 24.89 $\pm$ 0.21 & Hashimoto et al. 2019 \\
    611 & 25.31 $\pm$ 0.19 & Tripodi et al. 2022 \\
    871 & 14.63 $\pm$ 0.34 & Tripodi et al. 2022 \\
    1037 & 11.77 $\pm$ 0.12 & Tripodi et al. 2022 \\
    1052 & 11.05 $\pm$ 0.16 & Tripodi et al. 2022 \\
    1130 & 8.81 $\pm$ 0.13 & Tripodi et al. 2022 \\
    1139 & 7.73 $\pm$ 0.31 & Tripodi et al. 2022 \\
    1199 & 8.29 $\pm$ 0.63 & Wang et al. 2013 \\
    1959 & 1.63 $\pm$ 0.06 & Tripodi et al. 2022 \\
    2126 & 1.4 $\pm$ 0.02 & Tripodi et al. 2022 \\
    2194 & 1.29 $\pm$ 0.03 & Tripodi et al. 2022 \\
    3028 & 0.4 $\pm$ 0.05 & Wang et al. 2013 \\
    3276 & 0.29 $\pm$ 0.01 & Tripodi et al. 2022 \\
    \hline

\end{longtable}
}

\end{appendix}


\end{document}